\documentclass[a4paper,11pt]{article}
\pdfoutput=1 

\usepackage{jcappub} 

\usepackage[T1]{fontenc} 
\usepackage{siunitx}
\usepackage{color}
\usepackage{indentfirst}

\title{\boldmath Track length measurement of $^{19}$F$^+$ ions with the MIMAC Dark Matter directional detector prototype}

\author[a,1]{Y. Tao,\note{Corresponding author.}}
\author[b]{C. Beaufort,}
\author[a,d]{I. Moric,}
\author[c,a]{C. Tao,}
\author[b]{D. Santos,}
\author[b]{N. Sauzet,}
\author[b]{C. Couturier,}
\author[b]{O. Guillaudin,}
\author[b]{J.F. Muraz,}
\author[b]{F. Naraghi,}
\author[d,a]{N. Zhou}
\author[c]{and J. Busto}

\affiliation[a]{Tsinghua Center for Astrophysics, Department of Physics, Tsinghua University, \\Beijing 100084, China}
\affiliation[b]{Laboratoire de Physique Subatomique et de Cosmologie, Universit Grenoble-Alpes(UGA), CNRS/IN2P3, Institut Polytechnique de Grenoble, \\53, rue des Martyrs, Grenoble, France}
\affiliation[c]{Centre de Physique des Particules de Marseille, Aix-Marseille Universit\'e, CNRS/IN2P3, \\Marseille, France}
\affiliation[d]{INPAC and School of Physics and Astronomy, Shanghai Jiao Tong University, Shanghai Laboratory for Particle Physics and Cosmology, \\Shanghai 200240, China}

\emailAdd{taoy15@mails.tsinghua.edu.cn}

\abstract{Weakly Interacting Massive Particles (WIMPs) are one of the most preferred candidate for Dark Matter.
WIMPs should interact with the nuclei of detectors. 
If a robust signal is eventually observed in direct detection experiments, the best signature to confirm its Galactic origin would be the nuclear recoil track direction \cite{Spergel1988}.  
The MIMAC collaboration has developed a low pressure gas detector providing both the kinetic energy and three-dimensional track reconstruction of nuclear recoils.
In this paper we report the first ever observations of $^{19}$F nuclei tracks in a $5$ cm drift prototype MIMAC detector, in the low kinetic energy range ($6$-$26$ keV), using specially developed ion beam facilities.
We have measured the recoil track lengths and found significant differences between our measurements and standard simulations. 
In order to understand these differences, we have performed a series of complementary experiments and simulations to study the impact of the diffusion and eventual systematics. 
We show an unexpected dependence of the number of read-out corresponding to the track on the electric field applied to the \SI{512}{\micro\meter} gap of the Micromegas detector. 
We have introduced, based on the flash-ADC observable, corrections in order to reconstruct the physical 3D track length of the primary electron clouds proposing the physics behind these corrections.
We show that diffusion and space charge effects need to be taken into account to explain the differences between measurements and standard simulations.
These measurements and simulations may shed a new light on the high-gain TPC ionization signals in general and particularly at low energy.
}

\begin{document}
\maketitle
\flushbottom

\section{Introduction}
\label{sec:intro}

The Dark Matter (DM) hypothesis plays a central role in cosmology and galaxy formation. 
The most widely accepted DM particle candidate is the Weakly Interacting Massive Particle (WIMP). 
Since Goodman and Witten \cite{GW} have proposed to detect the nuclear recoils produced by WIMP elastic collisions on active volume detection nuclei, many DM detectors have been developed and operated. 
Within the next decade, we expect that large mass non-directional detectors will either observe a candidate DM signal, or reach the solar neutrino floor. 
Spergel \cite{Spergel1988} has proposed to use a directional DM detector to demonstrate the Galactic origin of an eventual DM signal. 

Recently, we have compared how different directional detectors, as anisotropic crystals, nuclear emulsions or low pressure gases may preserve the initial nuclear recoil direction information \cite{Camille17}.
The study has shown that TPCs at low pressure (50 mbar)  provide the best access to this information, in terms of measurable tracks and angular resolution. 
Projects such as DRIFT \cite{daw11} (USA, UK), D$^3$ \cite{Ross2014} (USA), DMTPC \cite{Deaconu17} (USA), NEWAGE \cite{miuchi07} (Japan), CYGNO \cite{CYGNOCollaboration2019} (Italy) and MIMAC \cite{Santos2010} (France-China) are trying to achieve directional detection with different techniques (see \cite{Mayet2016}, \cite{Ahlen2009} or \cite{battat16} for a review).
MIMAC is an international collaboration trying to define the best high-definition configuration (detectors, read-out, gas mixture) for a future large detector having the possibility to perform directional DM detection.
Besides, there is a world-wide collaboration called CYGNUS, aiming at developing a global network for directional DM search. 

This paper addresses the performance of a prototype of a MIMAC (MIcro-TPC MAtrix of Chambers) directional DM detector, a time projection chamber with a mixed gas at low pressure (50 mbar) which can measure a few keV 3D nuclear recoil tracks. 
It provides simultaneous measurements of the ionization energy and 3D direction information.
We report the first ever track length measurements at low nuclear recoil kinetic energies ($6$ to $26$ keV). 
The experimental setup, presented in Section~\ref{sec:experimental}, consists of a MIMAC chamber prototype coupled to an original ion beam facility. 
In Section~\ref{sec:recon} we explain how we define and reconstruct the nuclear recoil tracks and its 3D length, in terms of transverse and longitudinal projections (or widths and depths, respectively in the rest of the paper).
In Section~\ref{sec:sim} we compare the measured 3D track depths and widths with simulations. 
We discuss the meaning of our length measurements, the possible origin of differences between observations and simulations, and the measurements that allows us to better understand these differences in Section~\ref{sec:systematics}.

\section{Experimental Setup and Low-energy Facilities}
\label{sec:experimental}

\subsection{MIMAC detector and principle of operation}
\label{sec:detector}

The MIMAC detector is a matrix of micro-Time Projection Chamber (TPC) (\cite{Sauli1977}, \cite{Billard2012} and \cite{Riffard2016}) developed in a collaboration between LPSC (Grenoble) and IRFU (Saclay). 
A chamber of the MIMAC matrix uses a direct coupling of a pixellized Micromegas with a specially developed fast self-triggered electronics.

The MIMAC gas mixture optimized for DM search is 70$\%$ CF$_{4}$ + 28$\%$ CHF$_{3}$ + 2$\%$ i-C$_{4}$H$_{10}$ at a pressure of $50$ mbar. 
The combination of the chosen gas mixture and pressure provides the necessary conditions of  high gain and drift velocity of primary electrons (about \SI{22}{\micro\meter}/ns) in the chamber, for 3D reconstruction of a few keV nuclear recoil tracks \cite{Couturier2016}. 
$^{19}$F, being a light odd nucleus represents an interesting target for spin dependent interactions, for which low pressure DM gas detectors are still competitive.
The gas mixture can be changed to explore other nuclear targets, which is one advantage of a gaseous detector.

\begin{figure*}[htbp]
    \centering
    \includegraphics[width=0.9\textwidth]{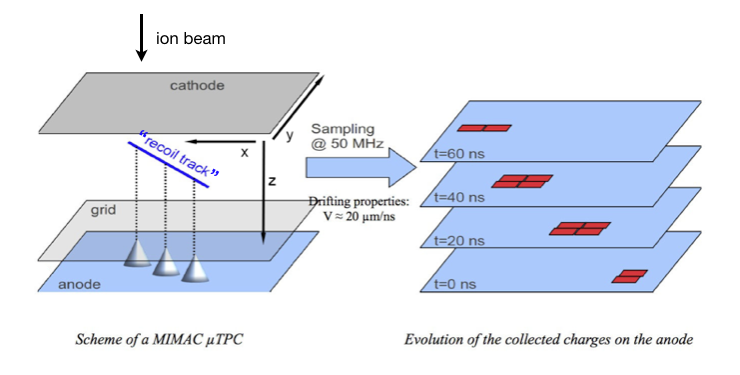}
    \caption{A simple scheme of a MIMAC detector chamber (left) and an example how sampling at 50 MHz is performed (right). 
    This configuration allows us to determine a 3D cloud of primary electrons and reconstruct the ion track. 
    The amplification gap in this experiment is \SI{512}{\micro\meter} with an electric field of over $11$ kV/cm.}
    \label{fig:Num1}
\end{figure*}

The nuclear recoil produced by an eventual elastic WIMP collision, or any ion injected in the detector, releases part of its kinetic energy in the form of ionization.
The primary electrons drift under an electric field of $150$ V/cm to the grid of a bulk Micromegas \cite{gio2006}, producing avalanches under the influence of a high electric field, greater than $10$ kV/cm in a thin \SI{512}{\micro\meter} amplification gap, as shown in Figure~\ref{fig:Num1}. 
 
The secondary electrons are then collected by the pixelated Micromegas anode, which contains strips of pixels in the $X$ and $Y$ directions (pitch of \SI{424.3}{\micro\meter}) with a total of $512$ channels ($256$ on each axis) over an area of $10.8\times10.8$ cm$^{2}$ \cite{iguaz2011}, providing a 2D readout. 
A strip is fired either along the $X$ or $Y$ direction when the collected charge is higher than a preset threshold. 
It is sampled at $50$ MHz ($20$ ns) by a self-triggered electronics system developed at LPSC \cite{Couturier2016}. 
The third spatial coordinate $Z$ is provided by the combination of the known primary electron drift velocity and the timing sampling. 
The electronics is based on a specially designed 64 channel MIMAC ASIC \cite{Richer2011} controlled by a data acquisition system \cite{Bourrion2011}.

The total ionization energy is measured by a charge pre-amplifier on the grid, by a Flash-ADC sampled also at $50$ MHz.
From the ionization energy value, it is possible to deduce the total recoil energy by taking into account the previously measured Ionization Quenching Factor (IQF) (\cite{Santos2010} and \cite{Billard2011}) corresponding to the fraction of the total kinetic energy released in ionization. 
This value depends on the nucleus, its kinetic energy, the gas mixture and gas pressure \cite{Couturier2016}. 
Existing models such as the Lindhard model \cite{Lind63}, and existing Monte Carlo simulations do not seem to provide a good description of experimental results at energies below $60$ keV \cite{RiffardThese}. 
That is why this IQF needs to be obtained experimentally for specifically defined configurations. 
IQF measurements for the MIMAC configuration were described in \cite{Gui2011} and \cite{RiffardThese}.

The aim of the ionization energy measurements and the 3D track reconstruction is to deduce the recoil kinetic energy and the direction of the initial scattered particle. 

Different experiments performed on LHI or COMIMAC (ion beam facilities described in Section~\ref{sec:beam-line}) coupled to the MIMAC chamber assessed the influence of gas and detector purity, optimizing the chamber electric field value and homogeneity, amplification gap thickness, anode pixel efficiency and event selection algorithms. 
This has provided invaluable experience in evaluating the impact of different detector properties on its performance in terms of track length of ions. 
LHI and COMIMAC are ion beam facilities described in Section~\ref{sec:beam-line}.

The main experiment reported here was performed at LPSC using the LHI beam line with a \SI{512}{\micro\meter} Micromegas bulk gap detector.  
The MIMAC prototype used was a $10.8 \times 10.8 \times 5$ cm$^{3}$ single chamber.
The grid voltage was set at $-570$ V and the cathode at $-1320$ V, while the anode was grounded. 
We also used a field cage in order to produce a uniform drift field.
The LHI beam line facility delivered ions of kinetic energies between $5$ keV and $25$ keV. 
It was coupled to the MIMAC chamber from the cathode side and ions were injected in the drift ($Z$-axis) direction at an angle of $\theta$ = 0$^{\circ}$ (same experimental configuration as \cite{Muraz2016}).
The final ion kinetic energies inside the chamber had an additional component due to the voltage applied on the cathode ($1.32$ kV) in order to have the electronic board grounded.

After entering the chamber, the injected ions immediately interact with the gas atoms and the produced primary electrons drift to the micromesh. 
There the electrons enter the gap with an intense electric field producing the avalanches. 
Secondary electrons are created and collected by the pixelated anode readout. 

The gain during the measurements can be estimated from the $^{55}$Fe peak ($5.9$ keV)  with a resolution of 19\% (FWHM) on the calibration spectrum. 
The total gain was estimated to  $2.2\times10^4$  considering the primary ionization energy to $38$ eV/pair.

\subsection{Low energy beam facilities: COMIMAC and LHI}
\label{sec:beam-line}

In order to measure the IQF and evaluate the performances of a MIMAC chamber detector, we have performed experiments on both COMIMAC and LHI facilities.

COMIMAC is a table-top ion beam facility developed at LPSC \cite{Muraz2016}. 
It delivers a continuous beam of electrons and mono-charged calibrated ions with a kinetic energy ranging up to $50$ keV. 
This facility is used to perform regular energy calibration using electrons and IQF measurements. 
COMIMAC uses a compacted $2.45$ GHz ($5$W) Electron Cyclotron Resonance (ECR) source called COMIC \cite{Muraz2016}. 
A Wien filter is used to make a charge-to-mass ratio ($q/m$) separation of ions and allows for their selection. 
The filter is a combination of a $0.36$ T magnetic field produced by permanent NdFeB magnets and a tunable $3.3$ kV/cm electric field in a perpendicular configuration. 

The LHI (Ligne exp\'erimentale \`a Haute Intensit\'e) is an experimental ion beam line based on a $8.5$ GHz ECR ion source coupled to a high resolution magnetic spectrometer. 
By applying a voltage on the plasma produced by the ECR ion source, the ions are extracted, collimated and sent to a high resolution magnetic spectrometer which separates the ion masses based on the $q/m$ factor over a trajectory which is an arc of circle with a radius of $\rho$= 0.7 m with $B\rho$ = $0.23$ T$\cdot$m. 
The LHI beam line produces ions with well defined energies and uncertainties on the kinetic energy at the level of $\cfrac{\Delta E_\text{kin}}{E_\text{kin}} = 1\%$. 
\vspace{0.5\baselineskip}

Both COMIMAC and LHI are coupled to the gas chamber via a \SI{1}{\micro\meter} diameter hole that was made by a laser on a \SI{13}{\micro\meter} thick stainless-steel foil.  
The hole interface coupled to a differential vacuum keeps a pressure independence between the beam line ($10^{-5}$ mbar) and the ionization chamber ($50$ mbar). 
The ions are thus injected in the direction of the beam line parallel to the drift field in the chamber.

\section{Track Reconstruction}
\label{sec:recon}

\begin{figure*}[htbp]
    \centering
    \includegraphics[width=0.6\textwidth]{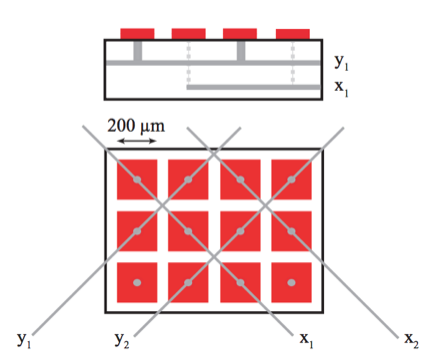}
    \caption{Readout electrode placed on the anode is segmented in $X$ and $Y$ direction strips providing 2D positional information for each event.}
    \label{fig:strips}
\end{figure*}

The secondary electrons created by the MIMAC Micromegas avalanche field reach the $X$ and $Y$ readout strips placed on the anode and provide the 2D positional information (Figure~\ref{fig:strips}). 
The collected secondary electrons by the Micromegas preserves the direction of the primary electrons.
The sampling of the anode every $20$ ns allows the reconstruction of a 3D cloud of primary electrons for each detected event.

\subsection{Track depth and width definitions}
\label{sec:definition} 

The ions delivered by LHI enter the chamber in the direction of the electron drift path along the $Z$-axis. 
We define the ion track depths as the projection of the primary electron cloud in the $Z$-direction:

\begin{equation}
    z_\text{max}-z_\text{min} = (t_\text{max}-t_\text{min}) \times V_\text{drift}, 
\end{equation}

where $V_\text{drift}$ is the electron drift velocity, and $t_\text{max} - t_\text{min}$ is the time difference between the first and last time signal of an ion event. 
The primary electron drift velocity was determined from MAGBOLTZ code \cite{Biagi1999} to be $V_\text{drift}$ = \SI{22.9}{\micro\meter}/ns (for an applied electric field of $150$ V/cm). 

Another available observable is the track width.  
We define it as the mean value of the number of strips triggered during a sampling interval on the $X$/$Y$ readout. 

\begin{figure*}[htbp]
    \centering
    \includegraphics[width=0.85\textwidth]{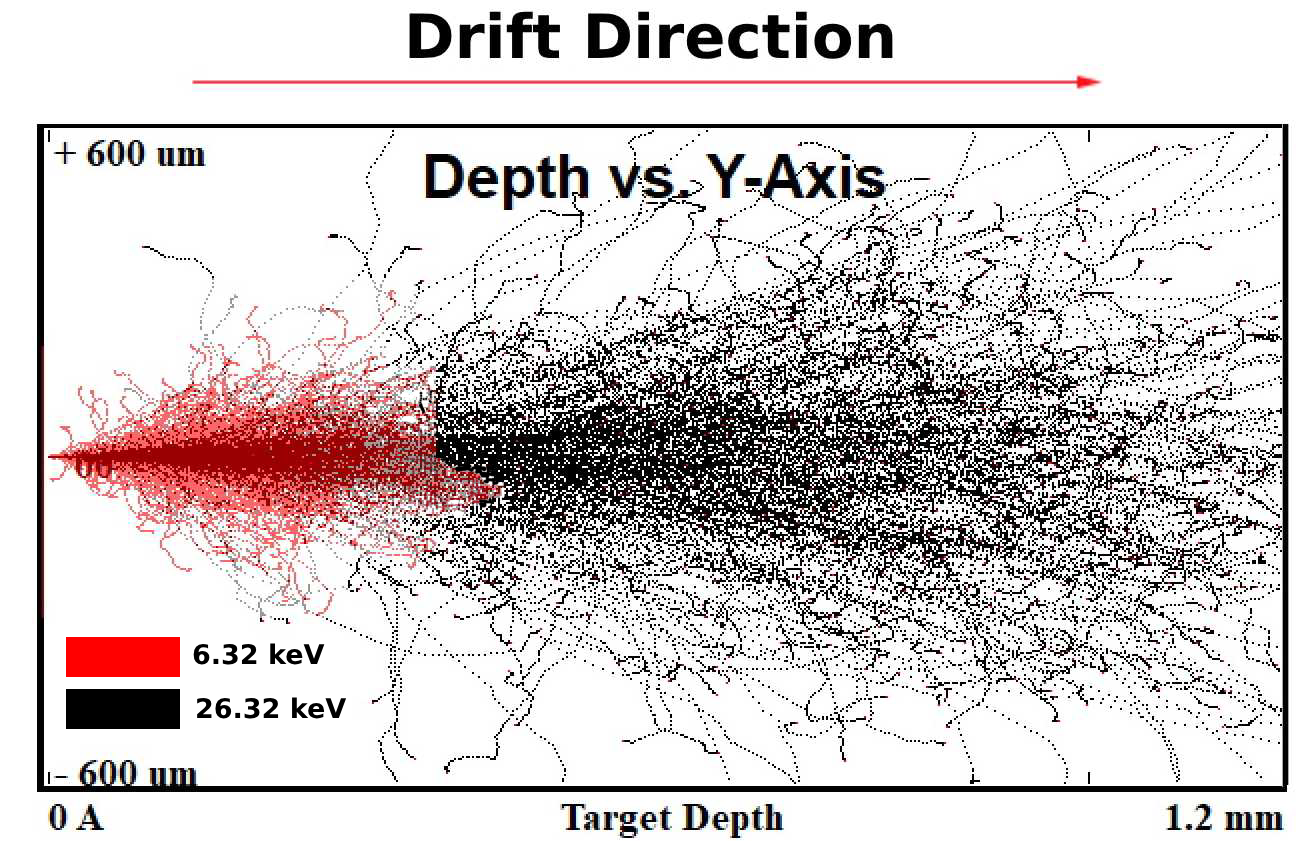}
    \caption{Taken from a SRIM simulation, this image shows how an ion path is deviated due to interactions with gas for ions of kinetic energy of 6.32 keV (in red) and 26.32 keV (in black). 
    The left vertical axis shows the position of the cathode, while the horizontal axis is the ion track depth. 
    The red arrow shows the drift direction of primary electrons. 
    The detected cloud of primary electrons therefore reflects not only the limitation of the detector to discern the initial track direction, but mostly the non-linear energy loss and multiple small-angle scattering of ions.}
    \label{fig:scattering}
\end{figure*}

The observed tracks are only a few mm long at such low energies. 
Figure~\ref{fig:scattering} shows an example provided by the SRIM simulation for $^{19}$F ions with kinetic energy of $6.3$ keV and $26.3$ keV, respectively. 

After the ionization in the active volume of the gas chamber, primary electrons have kinetic energies of the order of a few eV. 
This energy will quickly be lost because of multiple interactions with the gas molecules leading to thermalization and recombination \cite{Sauli1977}. 
By applying an electric field, the electrons drift towards the anode and their 3D Gaussian distribution $n(x,y,z;t)$ can be described as:

\begin{equation}
    n(x,y,z;t) = \frac{n_{0}}{\sqrt{8\pi^{3}}} \times \frac{e^{-(x^{2}+y^{2})/4D_\text{t}t}}{\sqrt{4D^{2}_\text{t}t^{2}}}\times \frac{e^{-z^{2}/4D_\text{l}t}}{\sqrt{2D_\text{l}t}}  
\end{equation}

where ${D_\text{t}}$ and ${D_\text{l}}$ are the transverse ($X$/$Y$) and longitudinal ($Z$) diffusion coefficients, respectively. 

Primary electrons experience transverse and longitudinal diffusion inside the gas chamber leading to longer and wider reconstructed track depths and widths, with the following standard deviations \cite{BillardThese}:

\begin{equation}
    \sigma_\text{t} = \widetilde{D}_\text{t}\sqrt{L_\text{d}} \textrm{\quad and\quad }
    \sigma_\text{l} = \widetilde{D}_\text{l}\sqrt{L_\text{d}},
\end{equation}

where $L_\text{d}$ is the electron drift distance, and $\widetilde{D}_\text{t/l} = \sqrt{2D_\text{t/l}/V_\text{drift}}$. 
For the MIMAC setup, we use for this paper, $L_\text{d}$ = $4.7$ cm. 

$\widetilde{D}_\text{t}$ and $\widetilde{D}_\text{l}$ can be obtained with the MAGBOLTZ code. 
Diffusion depends on the type of gas and its pressure and on the drift electric field. 
At the drift electric field ($150$ V/cm) applied in the MIMAC chamber, the MAGBOLTZ simulation predicts the following transverse and longitudinal diffusion :

\begin{equation}
  \begin{cases}
    \widetilde{D_\text{t}} = \SI{253.1}{\micro\meter} / \sqrt{\textrm{cm}} \\  
  \\[2pt]
    \widetilde{D_\text{l}} = \SI{293.9}{\micro\meter} / \sqrt{\textrm{cm}}  
  \label{eq:diffusion}
  \end{cases}
\end{equation}

Once the primary electrons reach the grid, they enter the amplification region where the avalanche will be developed.
We first assume that the electric field ($\mathcal{O}(10\mathrm{kV}\cdot\mathrm{cm^{-1}})$) in the multiplication gap is  uniformly distributed, and avalanches take place in the entire multiplication gap.
The multiplication factor $M$ (the gain)  in the gap can be written, in such case, as

\begin{equation}
  M = e^{\alpha d}
\end{equation}

where $d = \SI{512}{\micro\meter}$, and $\alpha$ is the 1$^\text{st}$ Townsend coefficient which depends on several parameters, including the gap size, electric field, gas components and pressure.

\subsection{Results of track depth and width measurement}
\label{sec:result-measure}

\begin{figure*}[htbp]
    \centering
    \includegraphics[width=0.85\textwidth]{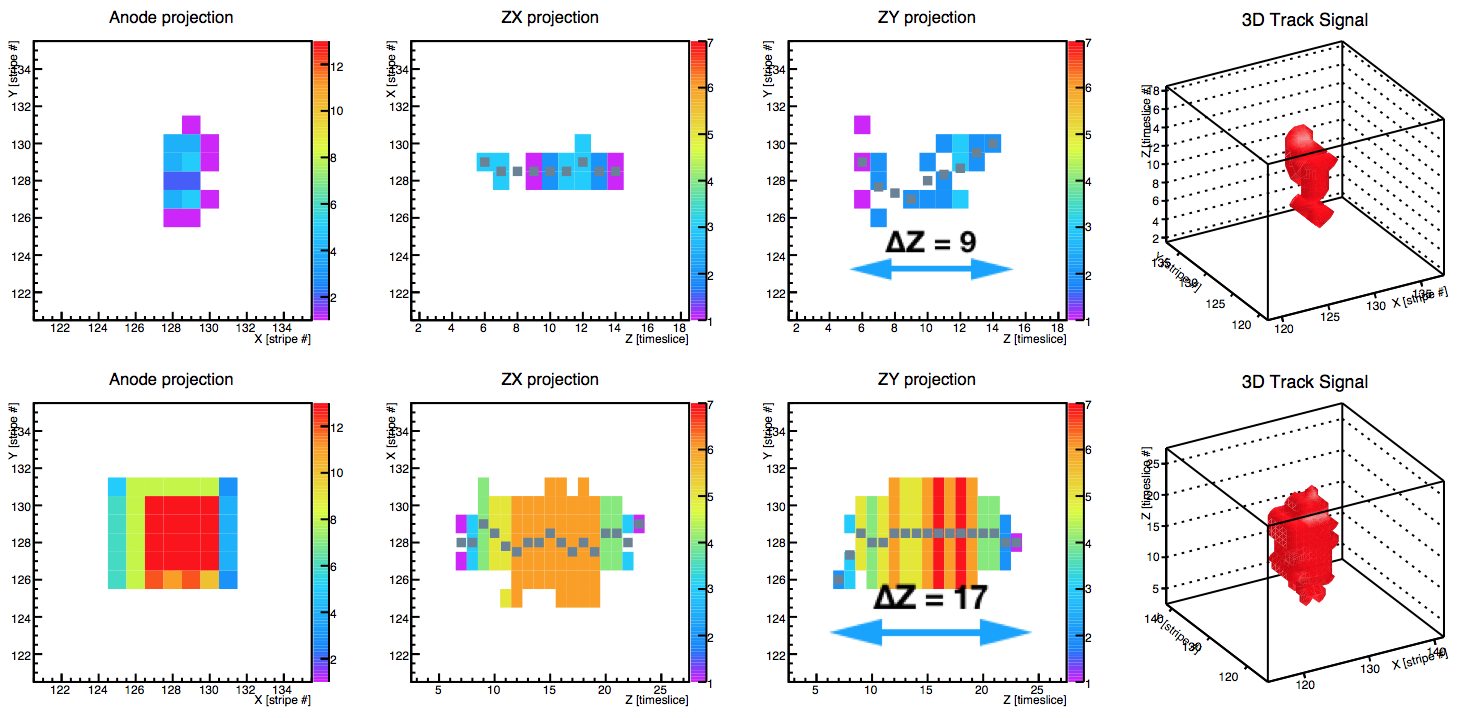}
    \caption{Measured $^{19}$F$^{+}$ ion tracks in $XY$, $ZX$, $ZY$ projection and 3D. 
    Top: a 6.3 keV kinetic energy example. 
    Bottom: an example of a $^{19}$F$^{+}$ track of 26.3 keV kinetic energy.}
    \label{fig:TrackExample}
\end{figure*}

The analysis was performed for $^{19}$F ions with kinetic energies of $6.3$ keV, $9.3$ keV, $11.3$ keV, $13.8$ keV, $16.3$ keV, $18.8$ keV, $21.3$ keV, $23.8$ keV and $26.3$ keV and with more than $1.8\times 10^{4}$ final events for each energy. 
Figure~\ref{fig:TrackExample} shows examples of track trajectories in $XY$, $ZX$, $ZY$ projections and in 3D for ions with kinetic energies of $6.3$ keV and $26.3$ keV, respectively.

\begin{figure*}[htbp]
    \centering
    \includegraphics[width=\textwidth, trim=50 0 50 0, clip]{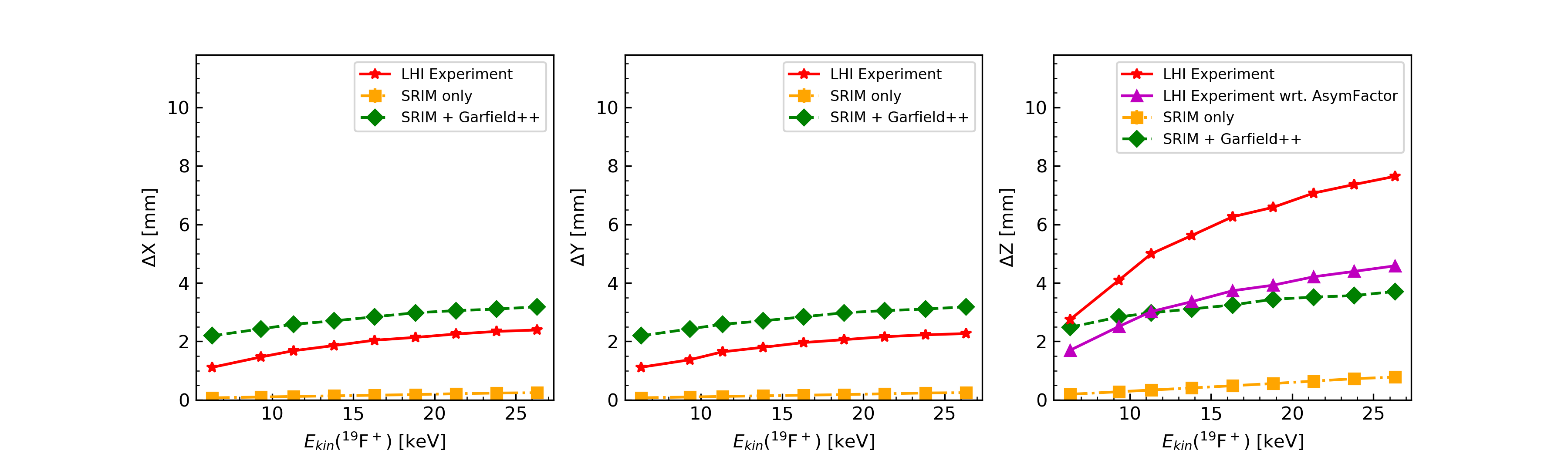}
    \caption{Comparison of ion track widths ($\Delta X$/$\Delta Y$) and depths ($\Delta Z$) at different energies between experiment (red stars) and Monte Carlo simulation (blue circles) combining SRIM and diffusion.
    The orange box is for SRIM only, the green diamond when diffusion and other effects are included using Garfield++ . 
    The magenta triangles are experimental measurements with an asymmetric factor correction.}
    \label{fig:DepthWidth}
\end{figure*}

The reconstructed average track depths and widths are shown in Figure \ref{fig:DepthWidth}, as a function of ion kinetic energies.
For the lowest ion kinetic energy of $6.3$ keV, a track of about $3$ mm depth and $1.5$ mm wide was measured. 
At the kinetic energy of $26.3$ keV, the ion tracks are showing a depth longer than $7$ mm, with an average width of $2.8$ mm.

\section{Comparison of Simulations with Measurements} 
\label{sec:sim}

Nuclear $^{19}$F track depth and width measurements have been compared to the simulations performed with the SRIM (Stopping and Range of Ions in Matter) code, a software allowing to calculate interactions of ions with matter (\cite{Ziegler2010} and \cite{Agostinelli2003}). 

It is based on a Monte Carlo simulation method, using the binary collision approximation with a random selection of the impact parameter of the next colliding ion. 
The inputs of SRIM include the type and initial energy of the ion, as well as the target definition and density. 
With these information, SRIM computes the three-dimensional distribution of the ions in the target and its parameters, such as penetration depth, its spread along the ion beam and perpendicular to it (called straggling); all target atom cascades in the target are followed in detail.
But the effects of drift and diffusion in the electric field of the chambers are not taken into account. 

We performed a set of simulations using Garfield++~\cite{Veenhof1998}, to estimate the diffusion and other effects during the drift.
Garfield++ can be interfaced with SRIM to get the ion energy loss in the medium and it calls MAGBOLTZ to obtain the gas properties. 
Billard \emph{et al.}\cite{Billard2014} have shown that MAGBOLTZ estimates for the primary electron velocity are similar to the measured ones in a pure CF$_4$ gas at $50$ mbar. 
Couturier \emph{et al.}\cite{Couturier2017} have reported measurements of the drift velocity performed by the MIMAC team with the same gas and pressure using the cathode signal
showing that in experimental conditions, the measured drift velocity suffers from a 12\% deviation compared to the MAGBOLTZ simulation. 
Since the drift velocity directly depends on the diffusion coefficients, we could expect similar deviations for the present depth and width measurements.
Garfield++ contains its own Monte Carlo code to transport the charged particles, which allows us to simulate the primary ionizations and the drift of primary electrons towards the grid, and the avalanche in a realistic situation. 
The simulated widths and depths are presented in Figure~\ref{fig:DepthWidth} (green diamonds) for 200 ion tracks per Fluorine kinetic energy bin, in comparison with the measurements.
The transport time of each primary electron depends on the electron path towards the grid and then suffers both from the longitudinal and the transverse diffusions.
Even in the presence of diffusion, the simulated depths are much shorter than the measurements.
This is not the case for the widths.

Our higher energy experiments which measured the track length of $^4$He recoil in MeV ionization energy range, show results consistent with simulations~\cite{Sauzet2019} in similar experimental configurations. 
There are also other efforts with different techniques and working conditions, which presented $^{19}$F recoil length measurements~\cite{DeaconuThese}, at energies higher than 30 keVee. 
This is the first time measurements have been performed in the low energy (6-26 keV) range with a \SI{512}{\micro\meter} Micromegas.

\section{Systematic effects}
\label{sec:systematics}
\begin{figure*}[htbp]	
    \centering
    \includegraphics[width=0.8\linewidth]{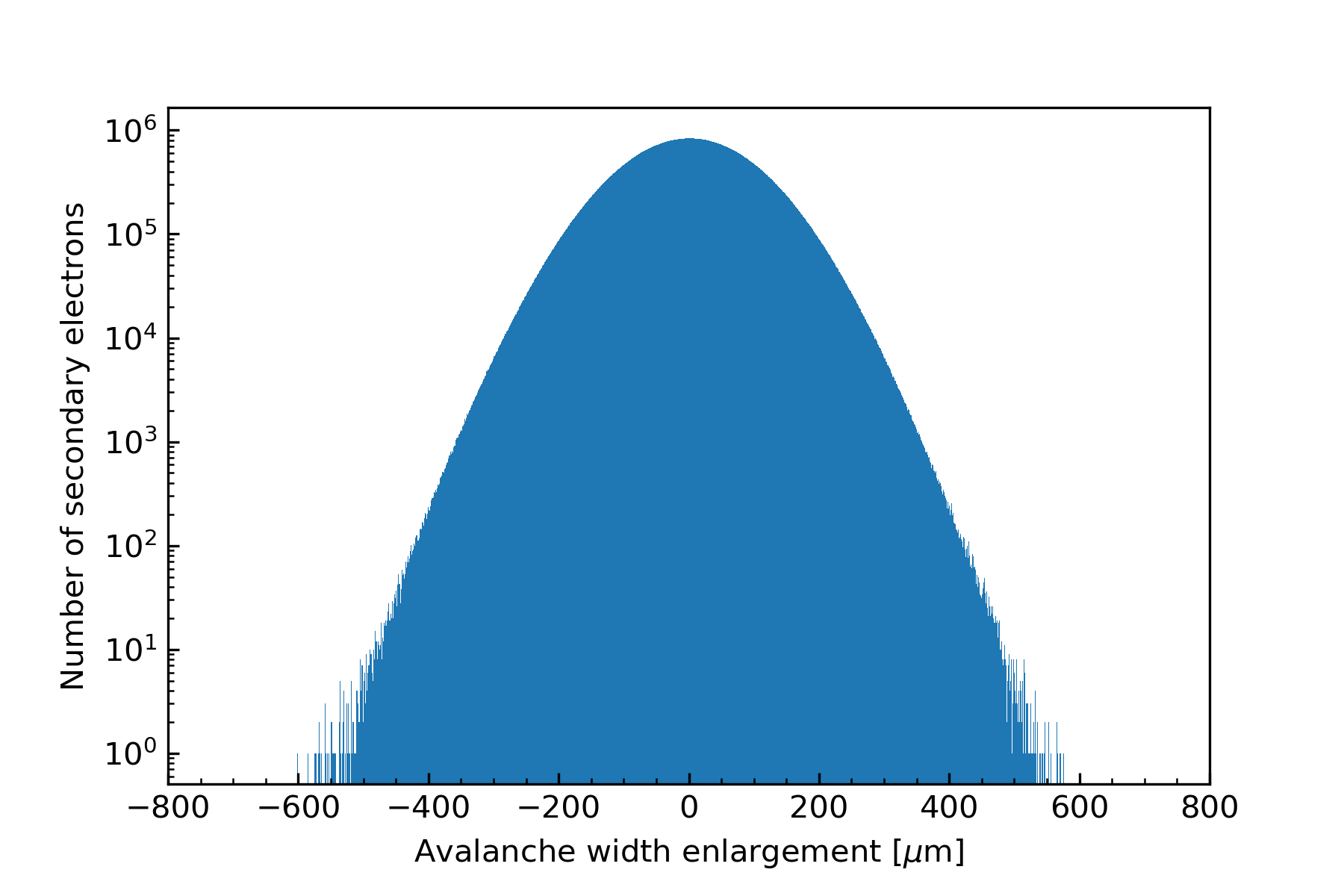}
    \caption{Additional contribution of secondary electrons to the width along $X$ direction, from Garfield++ simulation of 200 Fluorine events (10 keV).}
    \label{fig:width}
\end{figure*}

\begin{figure*}[htbp]
    \centering
    \includegraphics[width=0.8\textwidth]{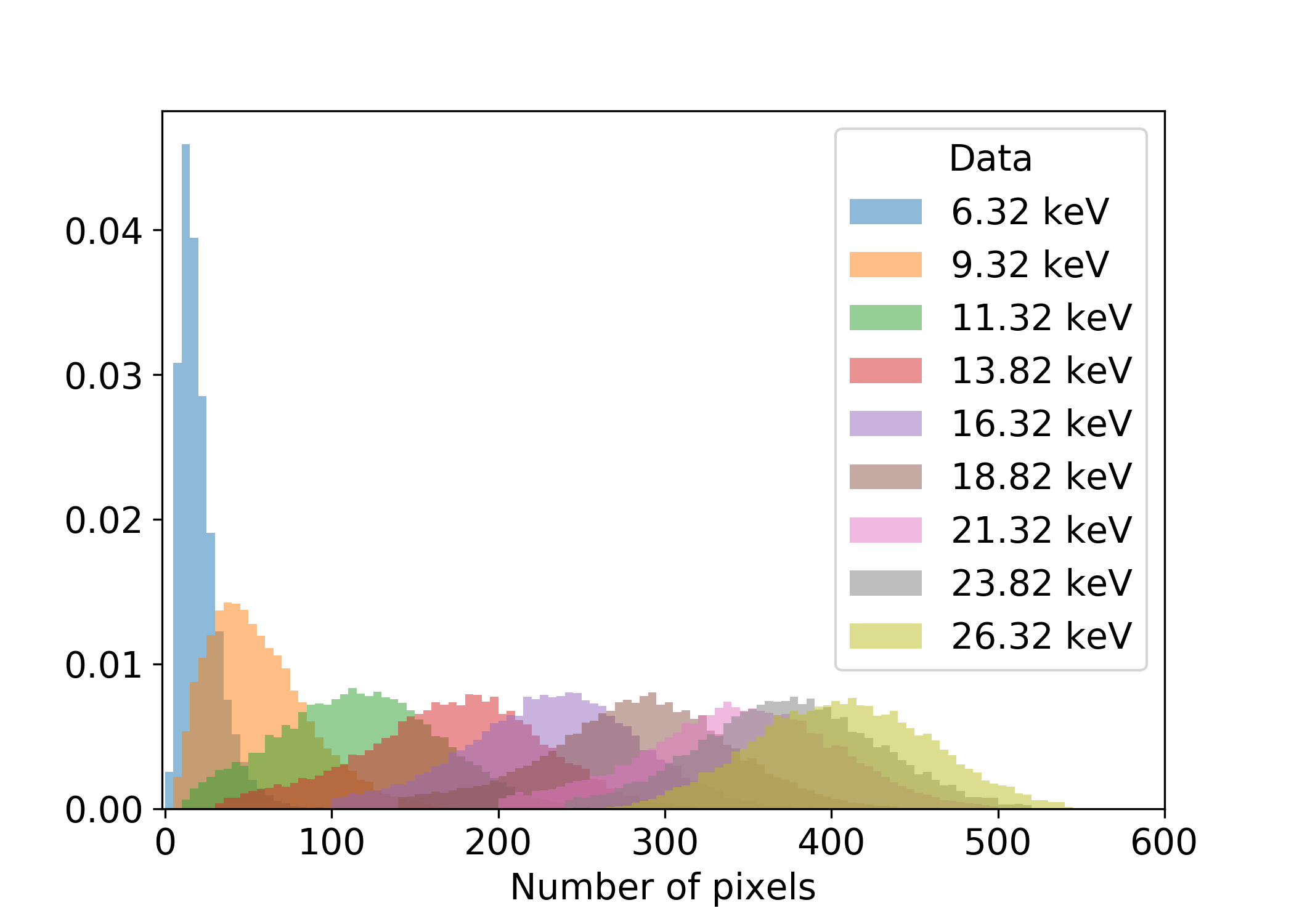}
    \caption{The number of pixels triggered for each kinetic energy of $^{19}$F$^{+}$ ion.}
    \label{fig:NPixels}
\end{figure*}

We have studied the following eventual systematic effects:

\begin{itemize}

  \item[-] If the detector anode strips have a lack of trigger efficiency, or the electron cloud is diluted, parts of the tracks would not be detected and this would result in shorter measured tracks but not longer tracks compared to the ones predicted by "SRIM + Garfield++". 

  \item[-] Garfield++ simulation (see Section~\ref{sec:sim-avala}) also shows a non-negligible contribution from the avalanche to the widths (see Figure~\ref{fig:width}), which means the detection efficiency is even lower than inferred from the left two plots in Figure~\ref{fig:DepthWidth}.

  \item[-] The spatial quantization or the strip/pixel  contributes no more than \SI{500}{\micro\meter} (see Figure~\ref{fig:strips}) to the uncertainty of width measurements in the transverse direction. 
A similar argument holds for depth measurements due to the fact that the product of drift velocity and sampling time is also less than \SI{500}{\micro\meter}.  

  \item[-] The number of pixels triggered for each energy is much larger than the requested threshold of 2 pixels and only at the lowest energy of $6$ keV (see Figure~\ref{fig:NPixels}), could we expect some bias. 
  
  \item[-] The diffusion during the drift significantly enlarges the primary electron cloud. 
  According to Garfield++ simulations, the diffusion dominates the measurements and influences some of the other systematic effects.
  The key point of a directional Dark Matter detector lies on the reconstruction of the direction of the incoming particle. 
  Analyzing the same dataset, we show in the companion paper~\cite{Tao2020a} that we obtain a better than 13$^\circ$ angular resolution for Fluorine ions with kinetic energies higher than 10 keV. 
  A detailed study leads us to conclude that such a precise angular resolution is achievable despite the diffusion. 
  This study shows that in fact  the 3D diffusion is the physical mechanism giving access to the details of the physical track.
  
  \item[-] The effect of space charge was measured and simulated for a Gas Electron Multiplier (GEM) based TPC~\cite{Bohmer}, so we studied this possible effect in the Micromegas amplification gap, as described below in Section~\ref{sec:sim-avala}.
  
\end{itemize}

\subsection{Avalanche multiplication}
\label{sec:sim-avala}
We had assumed so far the field in the amplification region to be uniform with negligible consequences.

Due to exponential growth, secondary electron-ion pairs are mainly created close to the anode plane and the secondary electrons drifting towards the anode in a short time of typically $\mathcal{O}(0.1\,\mathrm{ns})$. 
Compared with the sampling time of 20 ns this process gives a negligible contribution to the drift time. 
However, the positive ions drift through the gap with a drift velocity three to four orders of magnitude lower than the electron drift velocity~\cite{Yamashita}, which can induce some systematic effects. 
One possibility of such systematic effects could be that the slow ion flow back in the amplification gap leads to a local distortion of the electric field.
Subsequent primary electrons are entering the amplification region before the secondary ions have drifted back to the grid or have been neutralized. 
The number of backflow ions will increase all along the path of primary electrons in the amplification region, leading to a larger distortion of the electric field and results in an effective charge collection velocity that decreases along the drift direction in the gap.

\begin{figure*}[htbp]	
    \centering
    \includegraphics[width=0.8\linewidth]{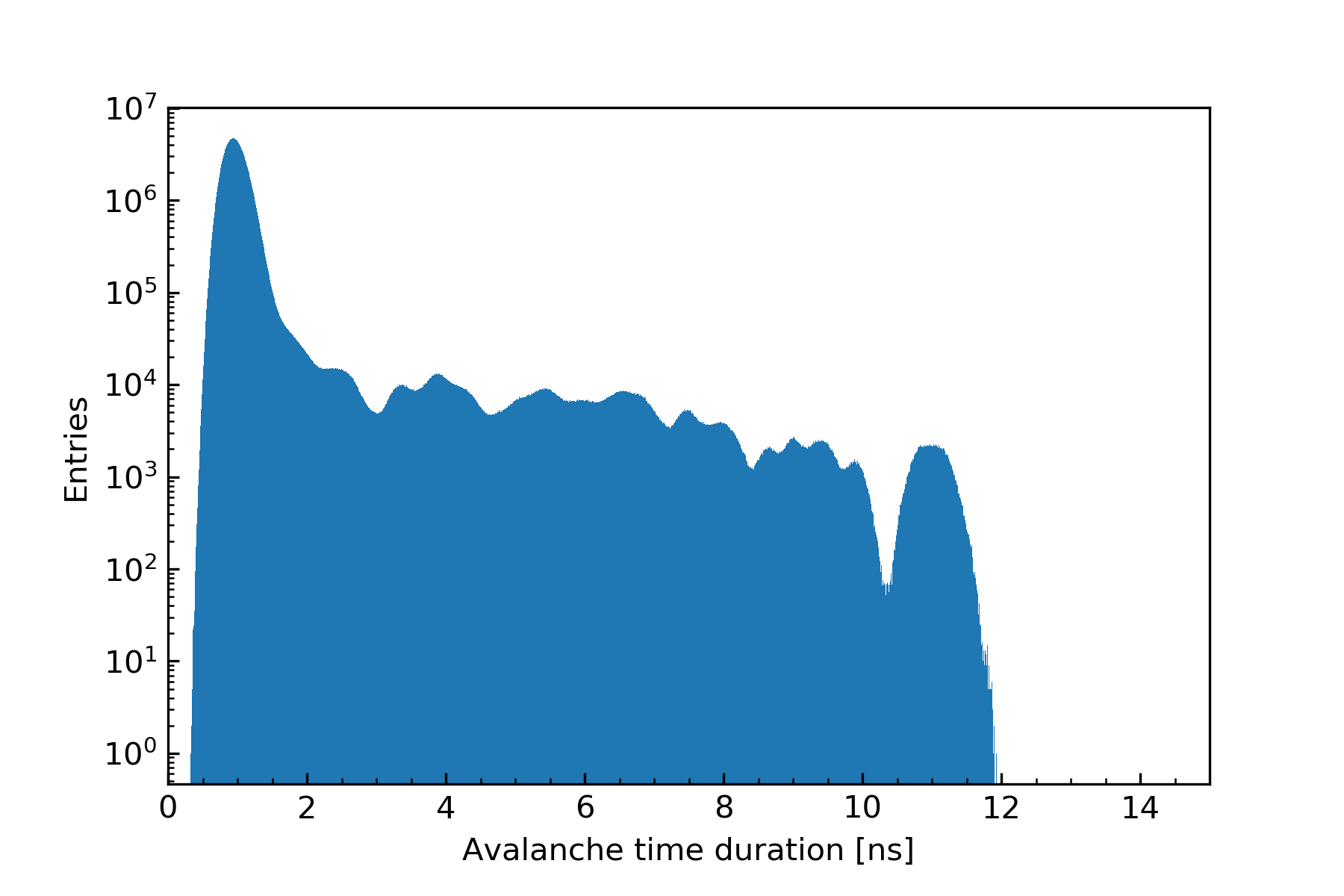}
    \caption{The avalanche time duration distribution of secondary electrons of 200 Fluorine events of 10 keV, simulated by Garfield++.}
    \label{fig:ava-dura}
\end{figure*}

Since MIMAC operates with a wide gap Micromegas (\SI{512}{\micro\meter}), we expect to be sensitive to this systematic effect producing a reduction of the effective drift velocity . 
We have simulated the avalanche generated by the primary electrons in the amplification region using Garfield++.
As shown in Figure~\ref{fig:ava-dura}, the typical drift time of secondary electrons is about a few ns. 
Thus, compared with the sampling time of 20 ns, the diffusion during the avalanche leads to a negligible contribution on the measured track depths.
So far, any other systematic effect, which would slow down secondary electrons, is not included in the Garfield++ simulations. 
The backflow of the secondary ions can distort local electric fields, leading to a decrease of effective drift velocity, and thus we measure more time slices than  predicted by simulations.

We performed another set of experiments with the COMIMAC facility to verify this hypothesis.
We sent electrons of $5\,\mathrm{keV}$ kinetic energy into the MIMAC prototype with a $5\,\mathrm{cm}$ drift chamber and a gas mixture of 50\% CHF$_3$ + 50\% i-C$_4$H$_{10}$ at $30\,\mathrm{mbar}$. 
We changed the gain by varying the amplification electric field in the range $[8.6\, ,\, 10.7]\,\mathrm{kV}\cdot\mathrm{cm}^{-1}$ while keeping a constant electric field at $150\,\mathrm{V}\cdot\mathrm{cm}^{-1}$ in the drift region.

Since the kinetic energy of incident electrons is fixed, their tracks with the same drift process are supposed to be the same all along the experiment.
We measured longer depths while increasing the gain, as shown in Figure~\ref{fig:asymFactor}. 
The experiments suggest a slowing down of the effective drift velocity as the density of secondary charges increases. 
Due to the slow ion backflow, this density depends not only on the gain but also on the energy of the incident particle.
Therefore, we conclude from these experiments that space charge effects in the amplification gap of Micromegas plays an important role in our track depth measurements.

\subsection{Correction factor for space charge effects}
\label{sec:correction}

\begin{figure*}[htbp]	
	\centering
	\includegraphics[width=0.9\linewidth]{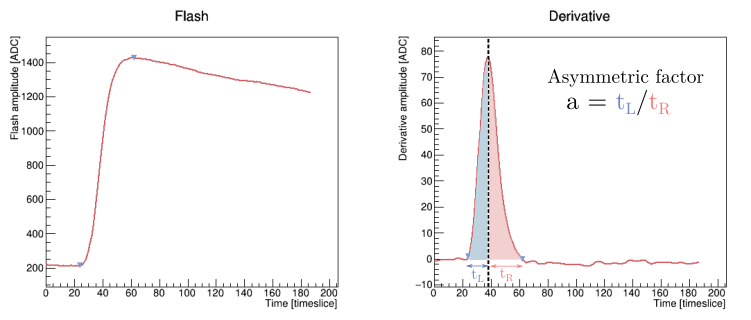}
	\caption{One example of a flash-ADC signal and its derivative curve in our measurement. 
	The asymmetry of the rise and decay time is shown on the right panel.}
	\label{fig:flashD}
\end{figure*}

The local electric field in the gap can be quite complicated, and it is not easy to model taking into account the distortion introduced by the ion backflow. 
The flash-ADC signal, integrating the charges arriving to the grid, provides a direct observation of this complex charge collection.
As the last charges to reach the grid will be more delayed, we expect the derivative of the flash-ADC signal to be enlarged due to the slower effective velocity. 
We define the "asymmetric factor'' as the ratio between the time duration of the falling part of the grid charge collection and the time duration of the rising part, see Figure~\ref{fig:flashD}. 
This asymmetric factor can be used to take into account the variation of the drift velocity. 

\begin{figure*}[htbp]	
	\centering
	\includegraphics[width=0.8\linewidth]{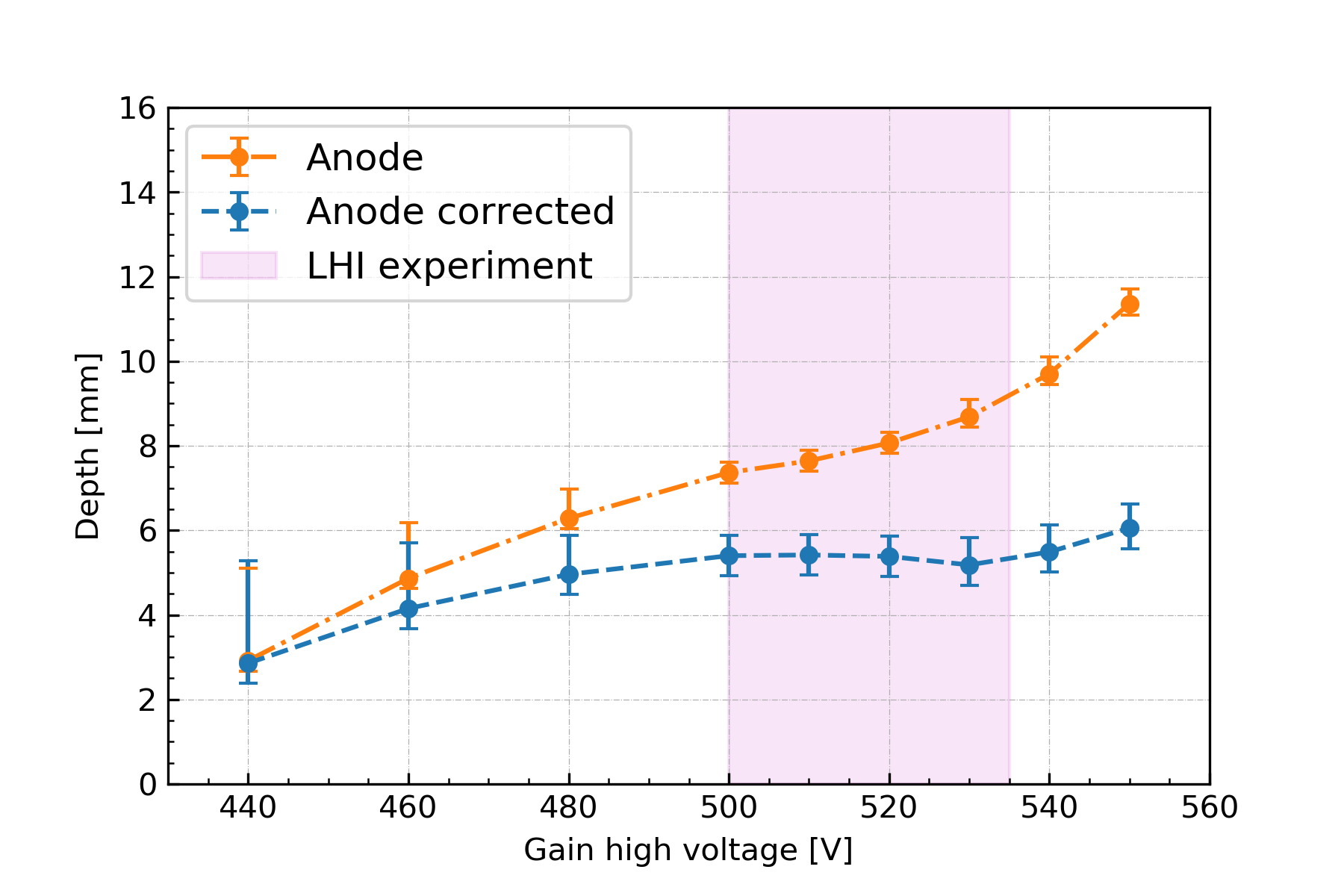}
	\caption{High-gain systematic effect and its empirical correction.}
	\label{fig:asymFactor}
\end{figure*}

Figure~\ref{fig:asymFactor} shows the depth measurements and the extracted contributions after we apply the asymmetric factor. 
The correction produces a quite constant depth when the gain increases, where the large uncertainties at low gain are explained by the anode strips lack of efficiency.
According to the values of the asymmetric factor, the densities of secondary charges in the LHI experiment are comparable with the ones of the shaded area of Figure~\ref{fig:asymFactor}. 
For this reason this correction can be applied to the LHI depth measurements.
Moreover, as shown by the magenta triangles in Figure~\ref{fig:DepthWidth}, the asymmetric factor applied on the depth measurements (shown as red stars) effectively gets rid of the contribution from systematic effects in the amplification gap and the extracted values are in agreement with the simulations performed.

However, since the high-gain systematic effect we referred here is not completely modeled or simulated in detail, the asymmetric factor can only be used as a first approximation and more  detailed calculations and simulations will be performed in the future.
This high-gain systematic effect should appear in other experiments with similar configurations and this work will be useful to understand the signals obtained. 
Our experiment using a \SI{512}{\micro\meter} Micromegas at high gain is the first measurement showing such effects.

\section{Conclusions}
\label{sec:conclusion}

MIMAC is currently the only collaboration that has presented 3D tracks for ions below 30 keV.
There are also other groups (\cite{Vahsen2014, Kohli2017}) making progress in 3D track reconstruction.
With a $10.8 \times 10.8 \times 5$ cm$^{3}$ prototype we present the reconstructed track length of $^{19}$F$^+$ ions - using the LHI and COMIMAC facilities to have well known kinetic energies.

Experimentally obtained track depths have been compared to simulations and were significantly longer than expected in the keV energy range using standard assumptions.
We found, sending the fluorine nuclei at the same kinetic energy, a variation of the number of anode read-outs on the avalanche electric field. 
This observation is interpreted as a space charge effect in the avalanche region, especially important at high gain. 
We used the observed asymmetry in the flash-ADC time signal to estimate a correction factor, which we use to infer the depth of the primary electron cloud.

The systematic effects presented in the paper influence significantly the track length measurements. 
Their impacts on the angular resolution are investigated in the companion paper~\cite{Tao2020a}.
Diffusion is often considered as having bad influence on track reconstruction. 
We have shown that it actually increases track lengths, allowing track detection and the reconstruction of the direction of the incoming particle if they are properly described. 
The directional information is a crucial point to measure the energy spectrum of neutronic fields via elastic scattering and a MIMAC detector with a 25 cm drift chamber has achieved such a detection~\cite{Sauzet2019, Tampon2018}.
Experimental measurements as presented in this paper will help understanding better gas detectors and their response to nuclear recoils for Dark Matter search.

\acknowledgments

Yi Tao, Igor Moric and Charling Tao thank Tsinghua University physics department and Department of Astronomy (DOA) and National Natural Science Foundation of China (NSFC11475205) for support. We acknowledge David Diez for developping the Garfield++ environment for the MIMAC team.

\bibliographystyle{JHEP}
\bibliography{references.bib}

\providecommand{\href}[2]{#2}\begingroup\raggedright\begin{thebibliography}{10}

\bibitem{Spergel1988}
D.~N. Spergel, \emph{{Motion of the Earth and the detection of weakly
  interacting massive particles}},
  \href{https://doi.org/10.1103/PhysRevD.37.1353}{\emph{Physical Review D}
  {\bfseries 37} (1988) 1353}.

\bibitem{GW}
M.~W. Goodman and E.~Witten, \emph{{Detectability of certain dark-matter
  candidates}}, \href{https://doi.org/10.1103/PhysRevD.31.3059}{\emph{Physical
  Review D} {\bfseries 31} (1985) 3059}.

\bibitem{Camille17}
C.~Couturier, J.~P. Zopounidis, N.~Sauzet, F.~Naraghi and D.~Santos,
  \emph{{Dark matter directional detection: comparison of the track direction
  determination}},
  \href{https://doi.org/10.1088/1475-7516/2017/01/027}{\emph{Journal of
  Cosmology and Astroparticle Physics} {\bfseries 2017} (2017) 27}.

\bibitem{daw11}
J.~B.~R. Battat, J.~Brack, E.~Daw, A.~Dorofeev, A.~C. Ezeribe, J.-L. Gauvreau
  et~al., \emph{{First background-free limit from a directional dark matter
  experiment: Results from a fully fiducialised DRIFT detector}},
  \href{https://doi.org/10.1016/J.DARK.2015.06.001}{\emph{Physics of the Dark
  Universe} {\bfseries 9-10} (2015) 1}.

\bibitem{Ross2014}
S.~Ross, \emph{{Recent Progress on D$^3$ -- The Directional Dark Matter
  Detector}},  \href{https://arxiv.org/abs/1402.0043}{{\ttfamily 1402.0043}}.

\bibitem{Deaconu17}
C.~Deaconu, M.~Leyton, R.~Corliss, G.~Druitt, R.~Eggleston, N.~Guerrero et~al.,
  \emph{{Measurement of the directional sensitivity of Dark Matter Time
  Projection Chamber detectors}},
  \href{https://doi.org/10.1103/PhysRevD.95.122002}{\emph{Physical Review D}
  {\bfseries 95} (2017) 122002}
  [\href{https://arxiv.org/abs/1705.05965}{{\ttfamily 1705.05965}}].

\bibitem{miuchi07}
K.~Nakamura, K.~Miuchi, T.~Tanimori, H.~Kubo, H.~Nishimura, J.~D. Parker
  et~al., \emph{{NEWAGE - Direction-sensitive Dark Matter Search Experiment}},
  \href{https://doi.org/10.1016/J.PHPRO.2014.12.091}{\emph{Physics Procedia}
  {\bfseries 61} (2015) 737}.

\bibitem{CYGNOCollaboration2019}
{CYGNO Collaboration}, \emph{{CYGNO: a CYGNUs Collaboration 1 m$^3$ Module with
  Optical Readout for Directional Dark Matter Search}},
  \href{https://arxiv.org/abs/1901.04190}{{\ttfamily 1901.04190}}.

\bibitem{Santos2010}
D.~Santos, G.~Bosson, J.~L. Bouly, O.~Bourrion, C.~Fourel, O.~Guillaudin
  et~al., \emph{{MIMAC: MIcro-tpc MAtrix of chambers for dark matter
  directional detection}},
  \href{https://doi.org/10.1088/1742-6596/469/1/012002}{\emph{Journal of
  Physics: Conference Series} {\bfseries 469} (2013) 12014}
  [\href{https://arxiv.org/abs/1311.0616}{{\ttfamily 1311.0616}}].

\bibitem{Mayet2016}
F.~Mayet, A.~M. Green, J.~B.~R. Battat, J.~Billard, N.~Bozorgnia, G.~B. Gelmini
  et~al., \emph{{A review of the discovery reach of directional Dark Matter
  detection}},  \href{https://arxiv.org/abs/1602.03781}{{\ttfamily
  1602.03781}}.

\bibitem{Ahlen2009}
S.~Ahlen, N.~Afshordi, J.~B.~R. Battat, J.~Billard, N.~Bozorgnia, S.~Burgos
  et~al., \emph{{The case for a directional dark matter detector and the status
  of current experimental efforts}},
  \href{https://doi.org/10.1142/S0217751X10048172}{\emph{International Journal
  of Modern Physics A} {\bfseries 25} (2009) 1}
  [\href{https://arxiv.org/abs/0911.0323}{{\ttfamily 0911.0323}}].

\bibitem{battat16}
J.~B.~R. Battat, I.~G. Irastorza, A.~Aleksandrov, T.~Asada, E.~Baracchini,
  J.~Billard et~al., \emph{{Readout technologies for directional WIMP Dark
  Matter detection}},
  \href{https://doi.org/10.1016/J.PHYSREP.2016.10.001}{\emph{Physics Reports}
  {\bfseries 662} (2016) 1}.

\bibitem{Sauli1977}
F.~Sauli, \emph{{Principles of operation of multiwire proportional and drift
  chambers." 1991. 79-188.}} CERN, 1991.

\bibitem{Billard2012}
J.~Billard, F.~Mayet, D.~Santos, F.~Mayet, D.~Santos, S.~F. Biagi et~al.,
  \emph{{Track reconstruction with MIMAC}},
  \href{https://doi.org/10.1051/eas/1253017}{\emph{EAS Publications Series}
  {\bfseries 53} (2012) 137}.

\bibitem{Riffard2016}
Q.~Riffard, D.~Santos, O.~Guillaudin, G.~Bosson, O.~Bourrion, J.~Bouvier
  et~al., \emph{{MIMAC low energy electron-recoil discrimination measured with
  fast neutrons}},
  \href{https://doi.org/10.1088/1748-0221/11/08/P08011}{\emph{Journal of
  Instrumentation} {\bfseries 11} (2016) P08011}
  [\href{https://arxiv.org/abs/1602.01738}{{\ttfamily 1602.01738}}].

\bibitem{Couturier2016}
C.~Couturier, O.~Guillaudin, F.~Naraghi, Q.~Riffard, D.~Santos, N.~Sauzet
  et~al., \emph{{Directional detection of Dark Matter with the MIcro-tpc MAtrix
  of Chambers}},  \href{https://arxiv.org/abs/1607.08765}{{\ttfamily
  1607.08765}}.

\bibitem{gio2006}
I.~Giomataris, R.~{De Oliveira}, S.~Andriamonje, S.~Aune, G.~Charpak, P.~Colas
  et~al., \emph{{Micromegas in a bulk}},
  \href{https://doi.org/10.1016/j.nima.2005.12.222}{\emph{Nuclear Instruments
  and Methods in Physics Research Section A: Accelerators, Spectrometers,
  Detectors and Associated Equipment} {\bfseries 560} (2006) 405}.

\bibitem{iguaz2011}
F.~J. Iguaz, D.~Atti{\'{e}}, D.~Calvet, P.~Colas, F.~Druillole, E.~Ferrer-Ribas
  et~al., \emph{{Micromegas detector developments for Dark Matter directional
  detection with MIMAC}},
  \href{https://doi.org/10.1088/1748-0221/6/07/P07002}{\emph{Journal of
  Instrumentation} {\bfseries 6} (2011) P07002}.

\bibitem{Richer2011}
J.~P. Richer, O.~Bourrion, G.~Bosson, O.~Guillaudin, F.~Mayet, D.~Santos
  et~al., \emph{{Development and validation of a 64 channel front end ASIC for
  3D directional detection for MIMAC}},
  \href{https://doi.org/10.1088/1748-0221/6/11/C11016}{\emph{Journal of
  Instrumentation} {\bfseries 6} (2011) C11016}.

\bibitem{Bourrion2011}
O.~Bourrion, G.~Bosson, C.~Grignon, J.~L. Bouly, J.~P. Richer, O.~Guillaudin
  et~al., \emph{{Data acquisition electronics and reconstruction software for
  real time 3D track reconstruction within the MIMAC project}},
  \href{https://doi.org/10.1088/1748-0221/6/11/C11003}{\emph{Journal of
  Instrumentation} {\bfseries 6} (2011) C11003}
  [\href{https://arxiv.org/abs/1110.4348}{{\ttfamily 1110.4348}}].

\bibitem{Billard2011}
J.~Billard, F.~Mayet and D.~Santos, \emph{{Assessing the discovery potential of
  directional detection of dark matter}},
  \href{https://doi.org/10.1103/PhysRevD.85.035006}{\emph{Physical Review D -
  Particles, Fields, Gravitation and Cosmology} {\bfseries 85} (2012) 35006}
  [\href{https://arxiv.org/abs/1110.6079}{{\ttfamily 1110.6079}}].

\bibitem{Lind63}
J.~Lindhard, V.~Nielsen, M.~Scharff and P.~V. Thomsen, \emph{{Integral
  equations governing radiation effects. (Notes on atomic collisions, III)}},
  {\emph{Kgl. Danske Videnskab. Selskab Mat.-fys. Medd.} {\bfseries 33} (1963)
  1}.

\bibitem{RiffardThese}
Q.~Riffard, \emph{{To cite this version : D{\'{e}}tection directionnelle de
  mati{\`{e}}re sombre avec MIMAC}}, Ph.D. thesis, Universit{\'{e}} of
  Grenoble, 2016.

\bibitem{Gui2011}
O.~Guillaudin, J.~Billard, G.~Bosson, O.~Bourrion, T.~Lamy, F.~Mayet et~al.,
  \emph{{Quenching factor measurement in low pressure gas detector for
  directional dark matter search}},
  \href{https://doi.org/10.1051/eas/1253015}{\emph{EAS Publications Series}
  {\bfseries 53} (2012) 119} [\href{https://arxiv.org/abs/1110.2042}{{\ttfamily
  1110.2042}}].

\bibitem{Muraz2016}
J.~F. Muraz, J.~M{\'{e}}dard, C.~Couturier, C.~Fourrel, O.~Guillaudin, T.~Lamy
  et~al., \emph{{A table-top ion and electron beam facility for ionization
  quenching measurement and gas detector calibration}},
  \href{https://doi.org/10.1016/j.nima.2016.06.107}{\emph{Nuclear Instruments
  and Methods in Physics Research, Section A: Accelerators, Spectrometers,
  Detectors and Associated Equipment} {\bfseries 832} (2016) 214}.

\bibitem{Biagi1999}
S.~F. Biagi, \emph{{Monte Carlo simulation of electron drift and diffusion in
  counting gases under the influence of electric and magnetic fields}},
  \href{https://doi.org/10.1016/S0168-9002(98)01233-9}{\emph{Nuclear
  Instruments and Methods in Physics Research Section A: Accelerators,
  Spectrometers, Detectors and Associated Equipment} {\bfseries 421} (1999)
  234}.

\bibitem{BillardThese}
J.~Billard, \emph{{D{\'{e}}tection directionnelle de mati{\`{e}}re sombre avec
  MIMAC}}, Ph.D. thesis, Universit{\'{e}} of Grenoble, 2012.

\bibitem{Ziegler2010}
J.~F. Ziegler, M.~D. Ziegler and J.~P. Biersack, \emph{{SRIM - The stopping and
  range of ions in matter (2010)}},
  \href{https://doi.org/10.1016/j.nimb.2010.02.091}{\emph{Nuclear Instruments
  and Methods in Physics Research, Section B: Beam Interactions with Materials
  and Atoms} {\bfseries 268} (2010) 1818}.

\bibitem{Agostinelli2003}
S.~Agostinelli, J.~Allison, K.~Amako, J.~Apostolakis, H.~Araujo, P.~Arce
  et~al., \emph{{Geant4--a simulation toolkit}},
  \href{https://doi.org/10.1016/S0168-9002(03)01368-8}{\emph{Nuclear
  Instruments and Methods in Physics Research Section A: Accelerators,
  Spectrometers, Detectors and Associated Equipment} {\bfseries 506} (2003)
  250}.

\bibitem{Veenhof1998}
R.~Veenhof, \emph{{GARFIELD, recent developments}},
  \href{https://doi.org/10.1016/S0168-9002(98)00851-1}{\emph{Nuclear
  Instruments and Methods in Physics Research, Section A: Accelerators,
  Spectrometers, Detectors and Associated Equipment} {\bfseries 419} (1998)
  726}.

\bibitem{Billard2014}
J.~Billard, F.~Mayet, G.~Bosson, O.~Bourrion, O.~Guillaudin, J.~Lamblin et~al.,
  \emph{{In situ measurement of the electron drift velocity for upcoming
  directional Dark Matter detectors}},
  \href{https://doi.org/10.1088/1748-0221/9/01/P01013}{\emph{Jinst} {\bfseries
  9} (2014) P01013} [\href{https://arxiv.org/abs/1305.2360}{{\ttfamily
  1305.2360}}].

\bibitem{Couturier2017}
C.~Couturier, Q.~Riffard, N.~Sauzet, O.~Guillaudin, F.~Naraghi and D.~Santos,
  \emph{{Cathode signal in a TPC directional detector: implementation and
  validation measuring the drift velocity}},
  \href{https://doi.org/10.1088/1748-0221/12/11/P11020}{\emph{Journal of
  Instrumentation} {\bfseries 12} (2017) P11020}.

\bibitem{Sauzet2019}
N.~Sauzet, D.~Santos, O.~Guillaudin, G.~Bosson, J.~Bouvier, T.~Descombes
  et~al., \emph{{Fast neutron spectroscopy from 1 MeV up to 15 MeV with
  Mimac-FastN, a mobile and directional fast neutron spectrometer}},
  \href{https://arxiv.org/abs/1906.03878}{{\ttfamily 1906.03878}}.

\bibitem{DeaconuThese}
C.~S. Deaconu, \emph{{A Model of the Directional Sensitivity of Low-Pressure
  CF4 Dark Matter Detectors}}, Ph.D. thesis, MASSACHUSETTS INSTITUTE OF
  TECHNOLOGY, 2015.

\bibitem{Tao2020a}
Y.~Tao, I.~Moric, C.~Beaufort, C.~Tao, D.~Santos, N.~Sauzet et~al.,
  \emph{{Angular resolution of a MIMAC Dark Matter directional detector
  prototype (will appear soon)}}, .

\bibitem{Bohmer}
F.~V. B{\"{o}}hmer, M.~Ball, S.~Dorheim, C.~H{\"{o}}ppner, B.~Ketzer,
  I.~Konorov et~al., \emph{{Simulation of Space-Charge Effects in an Ungated
  GEM-based TPC}},
  \href{https://doi.org/10.1016/j.nima.2013.04.020}{\emph{Nucl. Instrum. Meth.}
  {\bfseries A719} (2013) 101}.

\bibitem{Yamashita}
T.~Yamashita, H.~Kobayashi, A.~Konaka, H.~Kurashige, K.~Miyake, M.~Morii
  et~al., \emph{{Measurements of the electron drift velocity and positive-ion
  mobility for gases containing CF4}},
  \href{https://doi.org/10.1016/0168-9002(89)91445-9}{\emph{Nuclear Instruments
  and Methods in Physics Research Section A: Accelerators, Spectrometers,
  Detectors and Associated Equipment} {\bfseries 283} (1989) 709}.

\bibitem{Vahsen2014}
S.~E. Vahsen, M.~T. Hedges, I.~Jaegle, S.~J. Ross, I.~S. Seong, T.~N. Thorpe
  et~al., \emph{{3-D Tracking of Nuclear Recoils in a Miniature Time Projection
  Chamber}},  \href{https://arxiv.org/abs/1407.7013}{{\ttfamily 1407.7013}}.

\bibitem{Kohli2017}
M.~K{\"{o}}hli, K.~Desch, M.~Gruber, J.~Kaminski, F.~P. Schmidt and T.~Wagner,
  \emph{{Novel Neutron Detectors based on the Time Projection Method}},
  \href{https://arxiv.org/abs/1708.03544}{{\ttfamily 1708.03544}}.

\bibitem{Tampon2018}
B.~Tampon, \emph{Qualification exp{\'{e}}rimentale de la microTPC
  LNE-IRSN-MIMAC comme instrument de r{\'{e}}f{\'{e}}rence pour les mesures en
  {\'{e}}nergie et en fluence de champs neutronique entre 27keV et 6,5 MeV},
  Ph.D. thesis, 12, 2018.

\end{thebibliography}\endgroup

\end{document}